\documentclass[reprint,aps,pra,showpacs]{revtex4-1}

\usepackage{amsmath,amsfonts,amssymb,mathtools,bm,commath,graphicx,hyperref,units,float}

\usepackage{color}

\DeclareMathOperator{\Tr}{Tr}

\providecommand{\qint}[2]{\int\frac{d^{#2}\! #1}{(2\pi)^{#2}}}

\begin{document}
\preprint{APS/123-QED}

\title{Quantum interference phenomena in the Casimir effect}

\author{Andrew A. Allocca}
 \affiliation{Joint Quantum Institute and Condensed Matter Theory Center, Department of Physics, University of Maryland, College Park,
  Maryland 20742-4111, USA} 

\author{Justin H. Wilson}
 \affiliation{Joint Quantum Institute and Condensed Matter Theory Center, Department of Physics, University of Maryland, College Park,
  Maryland 20742-4111, USA}

\author{Victor Galitski}
 \affiliation{Joint Quantum Institute and Condensed Matter Theory Center, Department of Physics,\\ University of Maryland, College Park,
  Maryland 20742-4111, USA and\\ School of Physics, Monash University, Melbourne, Victoria 3800, Australia}

\date{\today}

\begin{abstract}
We propose a definitive test of whether plates involved in Casimir experiments should be modeled with ballistic or diffusive electrons--a prominent controversy highlighted by a number of conflicting experiments. The unambiguous test we propose is a measurement of the Casimir force between a disordered quasi-2D metallic plate and a three-dimensional metallic system at low temperatures, in which disorder-induced weak localization effects modify the well-known Drude result in an experimentally tunable way. We calculate the weak localization correction to the Casimir force as a function of magnetic field and temperature and demonstrate that the quantum interference suppression of the Casimir force is a strong, observable effect. The coexistence of weak localization suppression in electronic transport and Casimir pressure would lend credence to the Drude theory of the Casimir effect, while the lack of such correlation would indicate a fundamental problem with the existing theory. We also study mesoscopic disorder fluctuations in the Casimir effect and estimate the width of the distribution of Casmir energies due to disorder fluctuations. 
\end{abstract}


\maketitle

\section{Introduction}
The Casimir effect \cite{Casimir1948} is an experimentally accessible phenomenon which in most physical systems is theoretically calculated by modeling metallic plates with one of two models: the Drude or plasma model.
These simplified models describe the linear response of electrons in the plates to an electromagnetic field.
While the Drude model describes diffusive electrons subject to a random disorder potential, the plasma model describes ballistic electrons unhindered by disorder.
These two models typically provide similar predictions of the Casimir force as a function of plate separation, with the plasma model predicting a slightly stronger attraction than the Drude model.

\begin{figure}[b]
\centering
\includegraphics[width=0.5\columnwidth]{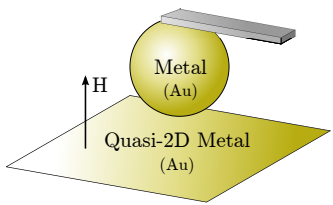}
\caption{\label{fig:experiment} (Color online) The geometry typically used in experimental measurements of the Casimir force is a gold coated sphere above a planar plate. Here we show the sphere suspended from a cantilever. We consider a lower plate of very thin metal with a weak applied perpendicular magnetic field.}
\end{figure}

Quantative results from many experiments \cite{Decca2005, *Decca2007, Castillo-Garza2013, Banishev2013} seem to favor the plasma model over a naive Drude model -- in some ways, arguably, the more physical of the two.
Many experiments attempt to account for the effect of electrostatic patch potentials in the plates, expecting the effect to be relevant for agreement with one model or the other.
Several of these \cite{Castillo-Garza2013,Banishev2013} find that the correction due to patches would make agreement with Drude worse while others \cite{Sushkov2011, Garcia-Sanchez2012} see agreement with the Drude model once the effect of patches is minimized.
There is recent theoretical work to account for the contribution of patch potentials in several experiments \cite{Behunin2012}, and while the results seem to weaken the case for the plasma model, the authors are careful to avoid claims that a definitive model for metallic plates has been found. 

\begin{figure}
\includegraphics[width=\columnwidth]{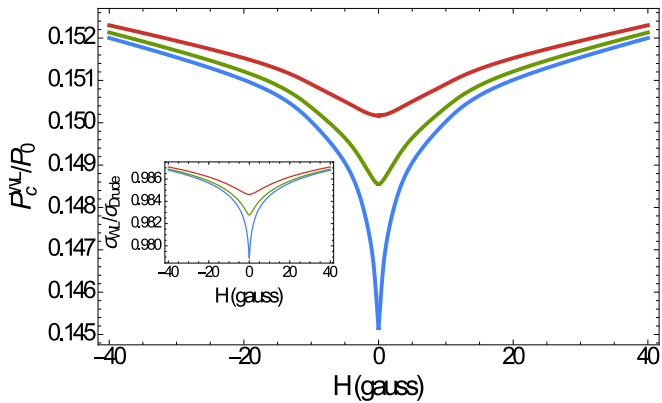}
\caption{\label{fig:mag} (Color online) The dependence of the Casimir pressure on the applied magnetic field between two disordered plates (one 3D and one 2D) at a separation of $a=\unit[250]{nm}$. The 2D plate is described by the Drude model with the weak localization correction. The force is normalized by the ideal conductor result and is plotted for three temperatures--$3$, $1$, and $\unit[0.1]{K}$, from top to bottom. The Casimir pressure is normalized by the ideal result, $P_0 = -\frac{\hbar c \pi^2}{240 a^4}$. The inset shows the conductivity of the 2D plate with WL correction as a function of the applied magnetic field, normalized by the uncorrected Drude conductivity, at the same three temperatures.}
\end{figure}

In this work, we provide a new way to experimentally test the validity of the diffusive electron model by tuning an external magnetic field (or temperature from a less practical standpoint) in a Casimir system with a two-dimensional plate. 
The proposed experiment would be the typical experiment seen in Fig.~\ref{fig:experiment} where the plate would be quasi-two-dimensional.
We find a dramatic change in the Casimir effect between Drude model plates due to weak localization, shown in Fig.~\ref{fig:mag}, that is just not seen with the plasma model.

Weak localization (WL) is a well known and greatly studied effect \cite{Hikami1980, Altshuler1980, Lee1985, Bergmann1984, Lin2002}, most easily observed in low-dimensional disordered systems at low temperatures where quantum interference logarithmically decreases the conductivity of a sample with decreasing temperature.
This is most easily understood via the simple Einstein relation, $\sigma = 2 e^2 \nu D$, where $\nu$ is the electronic density of states per spin in the material and $D$ is the diffusion constant.
Weak localization provides a quantum correction to the diffusion constant, $D \to D + \delta D$, that is strongly dependent on both temperature and an applied magnetic field at very low temperatures. 

A fundamental assumption of the WL effect is that electronic motion is diffusive in nature, and its contribution to conductivity is calculated as a correction to the Drude model. 
Therefore, any impact found on the nature of the Casimir effect due to WL would apply only to a diffusive model of metallic plates and not a ballistic model; a sensitive experimental test of the effects of WL on the Casimir effect would provide a clear indication of whether a diffusive picture of electronic motion correctly describes the physics of the electrons in the experiment.

The theory behind the use of a diffusive models relies upon performing an average over all possible realization of a disorder potential in the material.
However, if this disorder average is done at the level of linear response instead of on the Casimir energy itself, then all effects from, e.g., the nonuniform nature of physical disorder realizations are neglected.
While exact calculation of these neglected effects is impossible, it is possible to estimate whether ignoring them nonetheless gives a valid approximation to the Casimir energy, even if doing so causes discrepancies in other quantities, such as the entropy.
Indeed, an important issue regarding the use of the Drude model in the Lifshitz formula--a typical method of calculating the Casimir effect--is its apparent failure to satisfy the Nernst heat theorem \cite{Bezerra2002,Bezerra2004}, which there has been rigorous theoretical work to resolve \cite{Bordag2011, Ingold2014}.
Performing the disorder averaging at the level of the Casimir energy should automatically resolve this apparent violation since exact treatment of any specific disorder potential must satisfy the Nernst theorem. 

In this paper, we first provide an introduction to the relevant aspects of the Casimir effect and our calculational methods in Sec.~\ref{sec:Casimir}.
In Sec.~\ref{sec:WeakLoc} we then examine the correction to the Drude model that gives the weak localization effect, and in Sec.~\ref{sec:Results} we discuss the effect that this correction has on the Casimir force and provide estimates for the size of the effect.
In Sec.~\ref{sec:Fluctuations} we justify our use of the Drude model when considering disorder in spite of concerns regarding the disorder averaging procedure by examining the effect that fluctuations in disorder realizations would have on the Casimir energy and the ability to distinguish the Drude and plasma models.
Finally, in Sec.~\ref{sec:Conclusion} we summarize our results.

\section{Casimir Effect}\label{sec:Casimir}
To explore the effects of WL on the Casimir effect, we consider a system consisting of two flat parallel plates: one thick plate described by the Drude model and one two-dimensional plate described by the Drude model with an additional term giving the weak localization effect.
With an experimental setup in the typical plate-sphere geometry, as shown in Fig.~\ref{fig:experiment}, the gold layer on the sphere is thick enough to be most accurately described as a three dimensional material.
Since the effect of weak localization in 3D materials is much weaker than in 2D films, we only consider the WL effect in the 2D plate. 
In addition to this system of primary interest, we also consider the Casimir pressure between two plasma plates and between two Drude plates without weak localization for points of comparison. 
The latter of these will also give the expected behavior of the system including the 2D plate with WL correction at sufficiently high magnetic fields to completely suppress the WL effect ($H \gtrsim \unit[100]{gauss}$).
In these cases we consider the same geometry, with one thick plate and a parallel 2D plate. 
For a calculation of the Casimir pressure in these systems we start from the well-known Lifshitz equation for the Casimir energy density at finite temperature \cite{Klimchitskaya2009}, 
\begin{widetext}
\begin{equation} \label{eq:casimirenergy}
\mathcal{E}_c(a) = k_B T \sum_{\{\omega_n\}}{}^{'}  \qint{q}{2} \left[\ln\left(1-r_{TM}^{(1)}r_{TM}^{(2)} e^{-2 \sqrt{q^2 + \omega_n^2} a}\right) + \ln\left(1-r_{TE}^{(1)}r_{TE}^{(2)} e^{-2 \sqrt{q^2 + \omega_n^2} a}\right) \right],
\end{equation}
\end{widetext}
which can be obtained by expanding the free energy of the two plate and photon system.
In this expression, $\sum{}^{'}$ denotes a sum over positive Matsubara frequencies, counting the $n=0$ term with half weight.
The functions $r_{TM}^{(i)}$ and $r_{TE}^{(i)}$ are the reflection coefficients of plate $i$ for the two polarizations of light.
The subscript $TM$ refers to the polarization where the magnetic field is perpendicular to the plane of incidence, and similarly for $TE$ with the electric field.
The reflection coefficients depend on both $q$ and the Matsubara frequency $\omega_n = 2\pi n k_B T$, and may also depend on the other parameters of the system under consideration, such as the applied magnetic field $H$ and additional temperature dependence. 
The reflection coefficients can be written explicitly in terms of the dielectric functions of the plates, $\epsilon_i(i\omega_n)$, or alternatively in terms of the electromagnetic linear response functions of the plates, $\Pi_i(i\omega_n)$, which are related to the dielectric functions as
\begin{equation}\label{eq:dielectric}
\epsilon(i\omega_n) = 1 - \Pi(i\omega_n)/\omega_n^2.
\end{equation}
The reflection coefficients have the form
\begin{equation}\label{eq:2Dreflection}
\begin{gathered}
r_{TM}^{2D}(q,i\omega_n) = \frac{\Pi(i\omega_n)}{\Pi(i\omega_n) - \tfrac{2\omega_n^2}{q_\perp} } \\
r_{TE}^{2D}(q,i\omega_n) = \frac{\Pi(i\omega_n)}{\Pi(i\omega_n) - 2q_\perp}
\end{gathered}
\end{equation}
for two-dimensional plates, where we have defined $q_\perp = \sqrt{q^2 + \omega_n^2}$. 
For very thick three-dimensional metallic plates (thickness $d\to \infty$), the reflection coefficients have the form
\begin{equation} \label{eq:3Dreflection}
\begin{gathered}
r_{TM}^{3D}(q,i\omega_n) = -\frac{\sqrt{q_\perp^2 - \Pi(i\omega_n)} - q_\perp +\tfrac{q_\perp}{\omega_n^2}\Pi(i\omega_n)}{\sqrt{q_\perp^2 - \Pi(i\omega_n)} + q_\perp - \tfrac{q_\perp}{\omega_n^2}\Pi(i\omega_n)} \\
r_{TE}^{3D}(q,i\omega_n) = -\frac{\sqrt{q_\perp^2 - \Pi(i\omega_n)} - q_\perp}{\sqrt{q_\perp^2 - \Pi(i\omega_n)} + q_\perp}.
\end{gathered}
\end{equation}
The Casimir pressure is found from Eq.~\eqref{eq:casimirenergy} by taking its derivative with respect to plate separation, $a$.

\section{Weak Localization Correction} \label{sec:WeakLoc}
\begin{figure}
\includegraphics[width=\columnwidth]{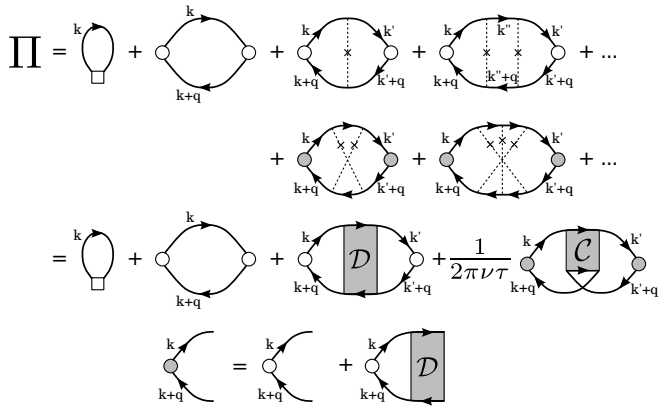}
\caption{\label{fig:correlation_diagrams} The diagrammatic expansion of $\Pi$ up to the leading correction to the Drude result. Solid lines represent disorder averaged electron Green's functions, dashed lines represent interactions with the disorder potential, the shaded regions represent diffusons (labeled with $\mathcal{D}$) or cooperons (labeled with $\mathcal{C}$) and the circles represent current vertices. The first three diagrams of the third line together give the Drude result, while the last term gives the leading correction. The last line defines the renormalized vertex.}
\end{figure}

The basic inputs into Eq.~\eqref{eq:casimirenergy} are the electromagnetic linear response functions for the two plates under consideration which contain all the electromagnetic properties necessary for a calculation of the Casimir effect. 
The response function of a non-interacting disordered electron gas can be represented diagrammatically as in Fig.~\ref{fig:correlation_diagrams}. 
The simplest approximation considers only the diagrams shown in the first line of the figure---those without impurity lines and all diagrams with impurity interaction ladders. 
In the long wavelength (i.e.\ local) limit, $\vec q \to 0$, these terms combine to give the response function $\Pi^{\text{Drude}} = -\frac{n e^2}{m}\frac{\omega_n}{\omega_n + 1/\tau}$, from which the well known Drude result for DC conductivity can be found, $\sigma^{\text{Drude}} = -\lim_{\omega_n \to 0} \Pi^{\text{Drude}}/\omega_n = \frac{n e^2 \tau}{m}$.
This approximation is valid for the one thick disordered plate we consider.

The leading correction to the Drude result, which cannot be ignored in an accurate treatment of the two dimensional plate at low temperatures, comes from diagrams with maximally crossed impurity lines, shown on the second line of Fig.~\ref{fig:correlation_diagrams}, which can be represented as a single diagram containing a cooperon. 
This approximation to the response function can be written as
\begin{equation}\label{eq:correlation}
\Pi = \Pi^{\text{Drude}} + \delta\Pi,
\end{equation}
where $\delta\Pi$ gives the WL correction.
An explicit calculation of $\delta\Pi$ in two dimensions, at low but finite temperature $T$, and with an external magnetic field $H$, gives the Hikami-Larkin-Nagaoka formula \cite{Hikami1980, Altshuler1980},
\begin{multline} \label{eq:correction}
\delta \Pi(i\omega_n; T,H) = \\ -\frac{e^2|\omega_n|}{2\pi^2\hbar} \left[\psi\left(\frac{1}{2}+\frac{\hbar}{4eH D\tau_\phi(T)}\right)-\ln\left(\frac{\hbar}{4eHD\tau}\right)\right],
\end{multline}
where $\psi$ is the digamma function, $D=\tfrac{v_F^2\tau}{2}$ is the diffusion constant, and $\tau_\phi$ is the electron dephasing time. 
This expression diverges logarithmically as $T\to 0$ with a sign opposite that of the Drude result, leading to a suppression of conductivity. 
It also has a very sensitive dependence on an applied magnetic field, becoming very small at moderate values of $H$ ($\sim\unit[100]{gauss}$) even at very low temperatures when the effect would be large in the absence of such a field.

In two dimensions the primary dephasing mechanism at very low temperatures is Nyquist electron-electron scattering, and the dephasing time is given by \cite{Altshuler1982}
\begin{equation}\label{eq:dephasing}
\frac{1}{\tau_\phi} = \frac{k_B T}{2\pi D \nu \hbar^2}\ln\left(\pi D\nu\hbar\right),
\end{equation}
where $\nu = m/2\pi \hbar^2$ is the density of states per spin at the Fermi level for a two dimensional system. 

\section{Results} \label{sec:Results}
We consider three systems of one thick plate and one 2D plate: both plasma plates, both Drude plates, and most importantly, a thick Drude plate with a 2D Drude plate including the weak localization correction term given in Eq.~\eqref{eq:correction}. 
In all of these systems, we fix the plate separation at $a=\unit[250]{nm}$ and calculate the Casimir pressure as a function of either an externally applied magnetic field or temperature, staying in the low temperature regime where Eq.~\eqref{eq:dephasing} is valid.
Additionally, we set the elastic mean free path of the electrons in disordered plates to be $l=\unit[15]{nm}$ and the Fermi energy and effective electron mass to be those of gold: $\epsilon_F=\unit[5.53]{eV}$ and $m^\ast = 1.10 m_0$ where $m_0$ is the free electron mass \cite{Ashcroft1976}.
We normalize all Casimir pressures by the ideal conductor result, $P_0 = -\frac{\hbar c \pi^2}{240 a^4}$, with $a=\unit[250]{nm}$.

\begin{figure}
\includegraphics[width=\columnwidth]{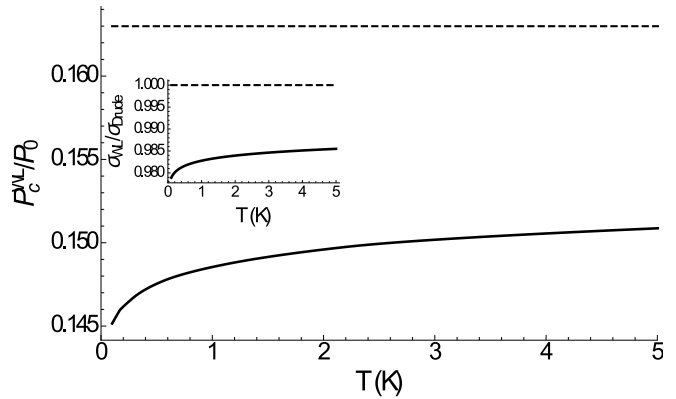}
\caption{\label{fig:temp} The dependence of the Casimir pressure on temperature between two Drude model plates (one 3D and one 2D) at a separation of $a=\unit[250]{nm}$. The force is normalized by the ideal conductor result, and there is no applied magnetic field. The solid line is obtained from including the WL correction in the 2D plate, and the dashed line is the result obtained if the effect of WL is ignored. The inset shows the dependence of the conductivity of the 2D plate as a function of temperature normalized by the uncorrected Drude model conductivity. The solid curve is obtained from the Drude model with WL correction and the dashed line at $1$ is for comparison to the uncorrected Drude model.}
\end{figure}

In addition to the disagreement on the magnitude of the effect between Drude and plasma models, we find that there is qualitatively different behavior between the two when accounting for the effect of weak localization. 
The Casimir pressure between plasma plates has no dependence on the strength of the applied magnetic field, at least for such weak fields as we consider here, and only a very weak dependence on temperature in this low temperature regime---the change of the normalized pressure from $\unit[10]{K}$ to $\unit[0.1]{K}$ is a decrease of $1.7\times 10^{-4}$.
In stark contrast, the Casimir pressure when considering a Drude plate with WL effects shows both a highly nontrivial dependence on even a weak applied magnetic field (at low temperatures), shown in Fig.~\ref{fig:mag}, and also a sharp decrease with decreasing temperature (with no applied magnetic field), shown in Fig.~\ref{fig:temp}. 
Both the temperature and magnetic field effects are expected when considering the Casimir pressure as a function of the conductivity of the plates. 
The sharp drop in the Casimir pressure with decreasing temperatures matches the drop in conductivity of the 2D plate obtained from theory, shown in the inset of Fig.~\ref{fig:temp}, and the strong dependence of the Casimir pressure on a weak magnetic field closely follows the dependence of the conductivity of the 2D plate as obtained from theory, shown in the inset of Fig.~\ref{fig:mag}, and seen in magnetoresistance experiments with 2D thin films \cite{Bergmann1984, Lin2002}.
Indeed, we find that applying a magnetic field of only $H=\unit[40]{gauss}$ perpendicular to the plates is enough to reduce the suppression of the pressure by approximately $40\%$.

At $T=\unit[0.1]{K}$ and $H=\unit[0]{gauss}$ we find that by correctly accounting for the effect of WL in the 2D plate the Casimir pressure is $11\%$ less than if the 2D plate were described by a simple Drude model without the WL correction.
At this temperature and magnetic field, the change in the Casimir pressure from including the WL correction is larger in magnitude than the difference in the Casimir pressures predicted by the plasma model and naive Drude model, i.e.,
\begin{equation} \label{eq:ratio}
\frac{P_c^\text{Drude}-P_c^\text{WL}}{P_c^\text{plasma}-P_c^\text{Drude}} =1.14 \text{  for } T=\unit[0.1]{K},
\end{equation}
so the effect is large enough to be measurable for a low enough temperature.

There are several ways to increase the size of the effect even beyond this, the most straightforward being to lower the temperature even further.
We also find that the effect can be increased by decreasing the electron mean free path, $l$, equivalent to increasing the impurity concentration, which can be seen by examining the dependence of Eq.~\eqref{eq:correction} on the mean free path, given partially through the dephasing time in Eq.~\eqref{eq:dephasing}.
When considering smaller values of $l$, however, one must be sure that the impurity concentration is still below the limit of complete Anderson localization, or else this model of diffusion breaks down.
Alternatively, when considering much larger values of $l$, which would make the effect smaller, one must be sure that the mean free path is much smaller than the sample dimension $L$ or else the model of a disordered system breaks down.
In an actual experimental system neither of these issues is likely to arise. 


Since a controlled smooth variation of temperature in Casimir effect experiments is almost an impossibility, especially at low temperatures where vibrational noise is difficult to remove due to the boiling of cryogenic liquids \cite{Decca2010}, an experimental test of the effects of weak localization on the Casimir effect could more easily be performed at fixed low temperature with varying magnetic field, looking for the effect shown in Fig.~\ref{fig:mag}.
Only weak magnetic fields would be necessary for such an experiment, as applying a magnetic field as weak as tens of gauss perpendicular to the plates would be enough to reduce the effect by a significant percentage.
We propose that an experimental test of these effects could be performed in the normal plate-sphere geometry with a very thin metallic film at a fixed separation, at a fixed low temperature, and with a varying weak magnetic field. 
While the exact numerical values the forces measured in this geometry are almost guaranteed to differ from the results we find, the general trends in the temperature and magnetic field dependence of the force are expected to remain. 

\section{Mesoscopic Disorder Fluctuations}\label{sec:Fluctuations}
When considering disordered systems one must determine when to perform averaging over disorder potential realizations.
Different realizations of the disorder potential will give different Casimir energies, and local fluctuations in the disorder will cause certain patches on each plate to vary in how attractive they are--very similar to the phenomenon of universal conductance fluctuations \cite{Altshuler1985, Lee1985a} (UCF)--leading to a self-averaging of the Casimir energy between two macroscopic plates. 
This argument would imply that instead of carrying out the averaging procedure on the linear response $\Pi$, which gives the Lifshitz formula with the Drude model, we should perform averaging over the entire Casimir energy itself.
Indeed, it is well known that using the Drude model in the Lifshitz formula can not be entirely correct, as it leads to a violation of the Nernst heat theorem \cite{Bezerra2002,Bezerra2004}, finding a non-zero entropy in the limit of zero temperature. 
In practice, however, it is not possible to consider an exact disorder potential or to perform the averaging procedure over the entire Lifshitz formula, and it is unknown if the simplification of using the disorder averaged linear response (i.e.\ the Drude model) in the Lifshitz formula is a legitimate approximation for Casimir force measurements despite the entropy problem.
Another way of phrasing this issue is that the approximation
\begin{equation}\label{eq:casimiraverage}
\left\langle\mathcal{E}_c[\Pi]\right\rangle = \mathcal{E}_c[\left\langle\Pi\right\rangle] + \delta\mathcal{E}_c \approx \mathcal{E}_c[\left\langle\Pi\right\rangle]
\end{equation}
certainly violates the Nernst theorem, but it is unclear if it nonetheless closely approximates the exact expression for the Casimir energy one would obtain if disorder were to be treated exactly or if averaging were done at the appropriate stage of the calculation.

Here we calculate what effect fluctuations from the average in any particular realization of a disorder potential have on the Casimir energy at low temperature, where conductance fluctuations are strongest.
We start from a microscopic version of the Lifshitz formula in position space,
\begin{equation} \label{eq:exactCasimir}
\mathcal{E}_c \left[ \Pi_1, \Pi_2 \right] = k_B T \sum_{\{\omega_n\}} {}^{'} \Tr \ln \left(1-M\right),
\end{equation}
where
\begin{equation*}
M =\!\! \int dr_1 dr_2 dr_3 \hat{\widetilde{\Pi}}_1(r,r_1) \hat{D}(r_1,r_2) \hat{\widetilde{\Pi}}_2(r_2,r_3') \hat{D}(r_3,r').
\end{equation*}
Here, $\hat{D}$ is the photon propagator, which in $M$ connect the screened response of one plate to the other, and $\hat{\widetilde{\Pi}}_i$ is the RPA screened electromagnetic linear response functions for plate $i$, schematically given by $(\hat{\bm{1}}-\hat{\Pi}_i \hat{D}(0))^{-1}\hat{\Pi}_i$, where $\hat{\Pi}$ is the unscreened linear response function and $\hat{D}(0)$ is the photon propagator along the plate.
From this point on we will consider plate 1 to be disordered with a particular disorder realization and for simplicity we will assume that plate 2 is homogeneous and not disordered. 
We will further take both plates to be two-dimensional.
For two-dimensional plates, the linear response function and photon propagator are $2\times 2$ matrices, with the components of $\Pi$ being proportional to the ac conductivity of the plates.
The trace, $\Tr$, is a generalized trace over both this matrix structure and the position labels of the function $M$.

\begin{figure}
\centering
\includegraphics[width=0.9\columnwidth]{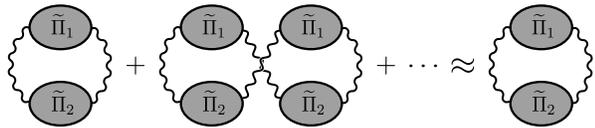}
\caption{The lowest order approximation to the Casimir energy given in Eq.~\eqref{eq:approximation}. The grey ovals represent the RPA screened linear response functions, and the wavy lines represent photon propagators.}
\label{fig:casimir1loop}
\end{figure}

We are interested in the case of metallic plates without a Hall effect so both response functions are proportional to the identity matrix. 
The photon propagator is diagonal as well. 
The matrix trace in Eq.~\eqref{eq:exactCasimir} becomes trivial, leaving us with a sum over photon polarizations. 
We make the further approximation that the Casimir energy is well described by just the first term obtained from expanding the logarithm,
\begin{widetext}
\begin{equation}\label{eq:approximation}
\mathcal{E}_c \left[ \Pi_1, \Pi_2 \right] \approx -k_B T \sum_{\{\omega_n\}}{}^{'} \int\prod_{i=1}^4 dr_i \!\!\! \sum_{X=TE,TM} \widetilde{\Pi}_1^X(r_4,r_1) D^{X}(r_1,r_2) \widetilde{\Pi}_2^X(r_2,r_3) D^{X}(r_3,r_4),
\end{equation} 
\end{widetext}
where now we label the two photon polarization with the superscript $X$.
Diagrammatically, this approximation can be represented as in Fig.~\ref{fig:casimir1loop}.

To ensure the following procedures are analytically tractable, we must make one further simplifying approximation.
We assume that the response function for plate 1, $\Pi_1$, which depends on an exact disorder realization, can be written as $\Pi_1 =\left\langle\Pi_1\right\rangle + \delta \Pi_1$, i.e.\ the exact response function can be written as the disorder averaged (Drude) response plus another small term to account for the particular disorder realization, here called $\delta\Pi_1$. 
(Note that this $\delta\Pi_1$ is distinct from the function of similar name given in Eq.~\eqref{eq:correlation} and Eq.~\eqref{eq:correction} which define a correction to the correlation function obtained after disorder averaging.)
With this notation, we now expand the RPA on plate 1 to first order in $\delta\Pi_1$ as,
\begin{multline}\label{eq:RPAapprox}
\widetilde{\Pi}_1 \approx \underbrace{\frac{\left\langle\Pi_1\right\rangle}{1-\left\langle\Pi_1\right\rangle D}}_{=\widetilde{\Pi}_1^\text{D}}  +\left[\frac{1}{1-\left\langle\Pi_1\right\rangle D} + \frac{\left\langle\Pi_1\right\rangle D}{\left(1-\left\langle\Pi_1\right\rangle D\right)^{2}}\right] \delta\Pi_1 \\
= \widetilde{\Pi}_1^D + \underbrace{\frac{1}{1-\left\langle\Pi_1\right\rangle D(0)}\left(1+\widetilde{\Pi}_1^D D(0)\right)}_{=\Gamma_1}\delta\Pi_1.
\end{multline}
This is the form of $\widetilde{\Pi}_1$ that is used in Eq.~\eqref{eq:approximation}.

\begin{figure}
\centering
\includegraphics[width=0.6\columnwidth]{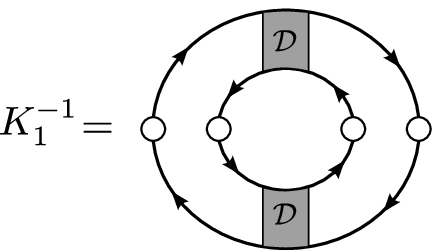}
\caption{The primary diagram giving the correlation of disorder fluctuations, as defined in Eq.~\eqref{eq:doubleBubble}. The components of the diagrams have the same meanings as given in Fig.~\ref{fig:correlation_diagrams}. Diagrams containing more diffusons are found either not to contribute to the correlator or are found to have a contribution $\mathcal{O}(1/\epsilon_F \tau)$ smaller.}
\label{fig:fluctuationdiagram}
\end{figure}

We now look to the probability distribution of the Casimir energy due to fluctuations in the disorder realization in plate 1, now contained entirely within the function $\delta\Pi_1$, which is given by
\begin{align} \label{eq:probIntegral}
\mathcal{P}_{\mathcal{E}_c}\left[ \mathcal{E} \right] &= \left\langle \delta\left(\mathcal{E}_c - \mathcal{E}\right)\right\rangle \nonumber \\
 &= \int \mathcal{D}\!\left(\delta\Pi_1\right) \int \frac{dx}{2\pi} \, e^{i x (\mathcal{E}_c -\mathcal{E})} \times  \\
 \times \exp&\left[-\frac{1}{2}\int\prod_{i=1}^4 dr_i \delta\Pi_1(r_1,r_2) K_1(r_1,\cdots,r_4) \delta\Pi_1(r_3,r_4)\right].\nonumber
\end{align}
This expression makes use of the disorder averaged correlator of two unaveraged response functions for the disordered plate, which can be written as,
\begin{equation} \label{eq:doubleBubble}
K_1^{-1}(r_1,r_2,r_3,r_4) = \left\langle \delta\Pi_1(r_1,r_2) \delta\Pi_1(r_3,r_4) \right\rangle,
\end{equation}
and is also given diagrammatically in Fig.~\ref{fig:fluctuationdiagram}.
The function $K_1^{-1}$ is very similar to the central object of interest considered in the context of universal conductance fluctuations \cite{Altshuler1985, Lee1985a, Rammer1991}, and it is calculated in the same manner.
It is related to the size of the fluctuations of the conductivity, $\delta\sigma^2$, (or equivalently in 2D, the conductance) in a similar way to how the linear response function $\Pi$ is related to the conductivity.
The only difference between the calculation of this function here and in the context of UCF is that the latter is primarily concerned with conduction of electrons through a system with attached leads, usually at zero temperature, while we consider a system with no leads at finite temperature.
As such, most of the qualitative properties of conductance fluctuations apply in our analysis of fluctuations in the Casimir energy as well, though the exact form of $K_1^{-1}$ differs by small numerical factors.
With this insight, we can already draw several conclusions about the nature of the distribution we will obtain from Eq.~\eqref{eq:probIntegral}.
Most importantly, for weak disorder we can expect fluctuations of the Casimir energy around the average value to be small since conductance fluctuations are small in good metals: $\delta\sigma^2/\sigma^2 \sim  1/(\epsilon_F\tau)^2$.
Additionally, we could expect the size of the fluctuations to be reduced by a factor of 2 if a magnetic field were applied to the sample.
This is because the diagram for $K_1^{-1}$ given in Fig.~\ref{fig:fluctuationdiagram} gives the same contribution at zero magnetic field if all diffusons are replaced with cooperons, but the cooperon contribution is suppressed in magnetic fields in the same way as the weak localization correction to the conductivity. 

\begin{figure}
\centering
\includegraphics[width=0.9\columnwidth]{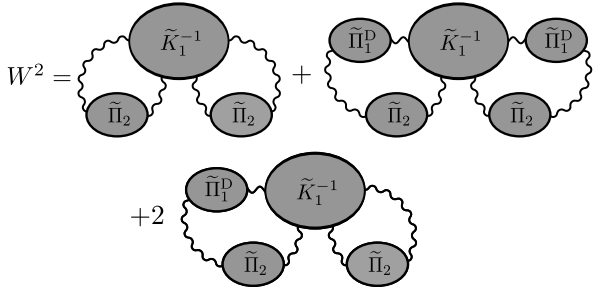}
\caption{The diagrams giving the width of the distribution, explicitly given in Eq.~\eqref{eq:width}.}
\label{fig:casimirfluct}
\end{figure}

In order to evaluate Eq.~\eqref{eq:probIntegral}, we perform a saddle point approximation on the functional integral over the disorder fluctuation, $\delta\Pi_1$, which after a straightforward calculation gives,
\begin{equation}\label{eq:distribution}
\mathcal{P}_{\mathcal{E}_c}[\mathcal{E}] = \frac{1}{\sqrt{2\pi W^2}}\exp\left[-\frac{(\mathcal{E}-\mathcal{E}_0^\text{Drude})^2}{2W^2}\right],
\end{equation}
where we have defined the quantities $\mathcal{E}_0^\text{Drude}$, the average, and $W$, the width of the energy distribution.
We find that the average energy is given by the same expression as in Eq.~\eqref{eq:approximation}, but with the substitution $\widetilde{\Pi}_1 \to \widetilde{\Pi}_1^\text{D}$, i.e.\ replacing the exact unaveraged response function $\Pi_1$ with the disorder averaged (Drude) response.
Therefore, the average $\mathcal{E}_0^\text{Drude}$ is simply an approximation of the exact Drude result.
Additionally, we find that the square of the width of the distribution can be written explicitly as,
\begin{widetext}
\begin{multline}\label{eq:width}
W^2 = (k_B T)^2 \!\!\!\!\! \sum_{\{\omega_n, \omega_{n'}\}}{}^{\!\!\!\!\!\!'} \int\prod_{i=1}^4 dr_i dr_i' \! \sum_{X,Y} D^X(r_1,r_2)\widetilde{\Pi}_2^X(r_2,r_3)D^X(r_3,r_4)\times \\
\times D^Y(r_1',r_2')\widetilde{\Pi}_2^Y(r_2',r_3')D^Y(r_3',r_4')\Gamma^X_1\Gamma^Y_1 K_1^{-1}(r_1,r_4,r_1'r_4'),
\end{multline}
\end{widetext}
which can be represented diagrammatically as in Fig.~\ref{fig:casimirfluct}. 
The multiple diagrams in this figure result from an expansion of the $\Gamma_1$ factors of Eq.~\eqref{eq:width}, 
\begin{align*}
\Gamma_1^X \Gamma_1^Y K_1^{-1} &= \frac{\left(1+\widetilde{\Pi}_1^\text{D} D^X(0)\right)\left(1+\widetilde{\Pi}_1^\text{D} D^Y(0)\right) K_1^{-1}}{\left(1-\left\langle\Pi_1\right\rangle D^X(0)\right)\left(1-\left\langle\Pi_1\right\rangle D^Y(0)\right)}\\
& \equiv \left(1+\widetilde{\Pi}_1^\text{D} D^X(0)\right)\left(1+\widetilde{\Pi}_1^\text{D} D^Y(0)\right) \widetilde{K}_1^{-1}.
\end{align*}

\begin{figure}
\centering
\includegraphics[width=\columnwidth]{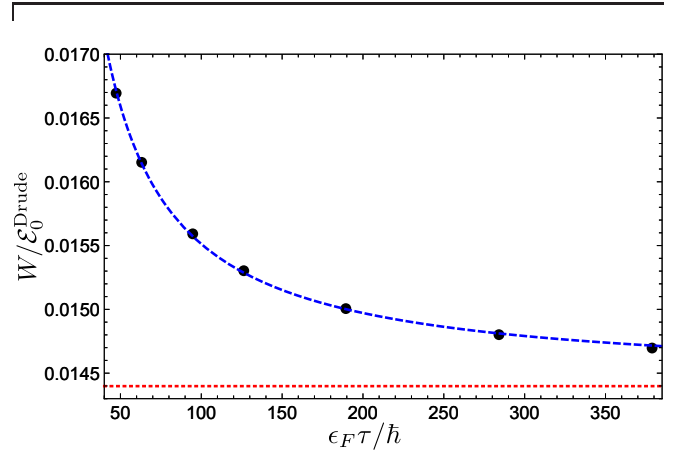}
\caption{(Color online) The fit of numerical data for the quantity $W/\mathcal{E}_0^\text{Drude}$ to the expected functional dependence given in Eq.~\eqref{eq:taudependence}. The black dots are the numerical data, the dashed blue line is the fit function, and the dotted red line is the asymptotic value $W/\mathcal{E}_0^\text{plasma}$, which has no dependence on $\tau$ in the leading approximation.}
\label{fig:WE0vstau}
\end{figure}

We compute these expressions numerically in the same way that we calculate the Casimir pressure in Sec.~\ref{sec:Results}. 
For both plates we use the Fermi energy and electron mass of gold, and we consider plate 1 to be disordered while plate 2 is a clean plasma plate.
We use the material parameters for gold in plate 2 so that in the limit of weakening disorder we are left with identical plasma plates. 
We vary the parameter $\tau$ to determine the dependence of $W$ and $E_0$, and the numerical results for their ratio are fit to the expected functional dependence, as shown in Fig.~\ref{fig:WE0vstau}.
In the parameter range we are interested in, we find the result,
\begin{equation} \label{eq:taudependence}
\frac{W}{\mathcal{E}_0^\text{Drude}} \approx \frac{W}{\mathcal{E}_c^\text{plasma}} + C_1\frac{\hbar}{\epsilon_F\tau}.
\end{equation}
In this expression $C_1 \approx 0.096$ is a distance independent constant and $\mathcal{E}_0^\text{plasma}$ is the Casimir energy between two clean plasma model plates calculated in the same approximation as $\mathcal{E}_0^\text{Drude}$, given in Eq.~\eqref{eq:approximation}.
In the same way, we also find this ratio's dependence on the distance between the plates.
It suffices to consider only the first term for this purpose, since the second term in Eq.~\eqref{eq:taudependence} has no dependence on $a$. 
We find,
\begin{equation}\label{eq:plasmadependence}
\frac{W}{\mathcal{E}_c^\text{plasma}} \approx C_2\sqrt{\frac{\hbar c}{\epsilon_F a}},
\end{equation}
where $C_2 \approx 0.038$ is another constant independent of both $\tau$ and $a$. 
The form of the disorder dependence in Eq.~\eqref{eq:taudependence} is expected since a weakening of disorder, interpreted as an increase of the scattering time $\tau$, will make a disordered plate more like a plasma plate.
Therefore, a very large scattering time should give a very good approximation to the plasma result.
Note, however, that the complete removal of disorder through the limit $\tau\to\infty$ has no physical meaning at this point in the calculation, since $W$ has already necessarily been calculated in the presence of disorder. 
We can see from these two expressions that the distribution will be relatively sharply peaked, in the sense that $W/\mathcal{E}_0^\text{Drude} \ll 1$, for plates that are not too close together and are in the disorder regime $1/\epsilon_F \tau \ll 1$, as we have considered thus far.

We can get a better understanding of how peaked the energy distribution is around its average value by comparing its width $W$ to a smaller relevant energy scale, $\mathcal{E}_0^\text{Drude} - \mathcal{E}_0^\text{plasma}$, by combining Eq.~\eqref{eq:taudependence} and Eq.~\eqref{eq:plasmadependence}.
We obtain,
\begin{equation}\label{eq:WvsEnergyDiff}
\frac{W}{|\mathcal{E}_0^\text{Drude}-\mathcal{E}_0^\text{plasma}|} \approx C_2\left(\sqrt{\frac{\hbar c}{\epsilon_F a}}+\frac{C_2}{C_1}\frac{c\tau}{a}\right).
\end{equation}
We see that the nature of the energy distribution Eq.~\eqref{eq:distribution} depends on the two dimensionless quantities $\hbar c/(\epsilon_F a)$ and $c\tau/a$.
Both of these dependencies can be understood intuitively.
The dependence on $\hbar c/(\epsilon_F a)$ can be understood as arising from the relevant photonic energy scale.
The most important photons are those with wavelength equal to twice the distance between the plates, and when this distance is large, these long wavelength photons are able to average over larger areas of the plates, reducing the effect of local fluctuations.
The dependence on $c\tau/a$ is similarly straightforward.
It is a comparison of two time scales: the impurity scattering time, $\tau$, and the time for photons to traverse the distance between the plates, $a/c$. 
When the ratio is small, electrons will have many impurity scattering events before interacting with a photon, so any effects due to impurities will be very important. 

There are several regimes we can now explore.
Here, we will always consider plates of the same material, so the Fermi energy is a fixed parameter and we can only vary $a$ and $\tau$.
First, if the plates are very close, meaning $\hbar c/(\epsilon_F a)$ is large, then the distribution is very wide regardless of the size of $\tau$.
Second, if $\tau$ is large compared to $a/c$, meaning that photons interact with any given electron many times between impurity scattering events, then the distribution is again very wide, regardless of the size of $\hbar c/(\epsilon_F a)$. 
The only regime in which the distribution is very sharply peaked is when both dimensionless parameters are small.
This requires that the plates are much farther apart than both length scales $\hbar c/\epsilon_F$ and $c \tau$, so that each electron undergoes many impurity scattering events between photon interactions and effect of disorder is more pronounced, but also so that long wavelength photons are most important, averaging out the disorder fluctuations. 

\begin{figure}[t]
\centering
\includegraphics[width=0.95\columnwidth]{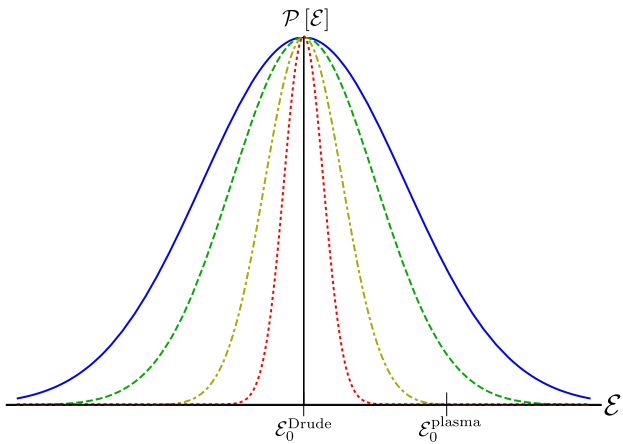}
\caption{(Color online) A plot of the distribution Eq.~\eqref{eq:distribution} given for several values of $a$ for a constant value of $\tau = \unit[4.5\times10^{-14}]{s}$, corresponding to $l=\unit[60]{nm}$. 
The values of $a$ are $250$ (solid blue), $400$ (dashed green), $800$ (dash-dotted yellow), and $\unit[1600]{nm}$ (dotted red). 
The average $\mathcal{E}_0^\text{Drude}$ and width $W$ are calculated numerically using Eq.~\eqref{eq:approximation} and Eq.~\eqref{eq:width}. 
The plots are scaled so the distributions are all the same height, and so $|\mathcal{E}_0^\text{Drude} - \mathcal{E}_0^\text{plasma}|$ is always the same width. 
Also indicated is the value of $\mathcal{E}_0^\text{plasma}$. 
One sees that as $a$ is increased the distribution becomes sharply peaked even compared to the small energy scale set by the difference from the plasma model.}
\label{fig:gaussians}
\end{figure}

Ultimately, this result means that for a given level of disorder, we can always go to large enough plate separations so that relatively small local fluctuations in the disorder potential of metallic plates are not likely to greatly affect the Casimir energy, as shown in Fig.~\ref{fig:gaussians}.
The difficulty here is that for large values of the inelastic scattering time $\tau$, the distance at which the distribution becomes very sharply peaked may be so large that the Casimir effect itself will become unmeasurably small.
We note now that the values of the parameters $\tau$ and $a$ used in Sec.~\ref{sec:WeakLoc} giving the results in Sec.~\ref{sec:Results} give a distribution very sharply peaked around its average, so we are justified in our use of the Drude model despite any of the stated concerns over the disorder averaging procedure. 

\section{Conclusion}\label{sec:Conclusion}
As we have shown, the weak localization correction to the Drude model at low temperatures may give the Casimir pressure a nontrivial dependence on both temperature and applied perpendicular magnetic field.
Moreover, we find that, for low enough temperatures, WL effects changes the Casimir pressure from the expected value without WL by an amount greater than the difference between the Drude and plasma model predictions. 
Since these effects are not applicable in a model of a 2D plate without disorder, i.e.\ the plasma model, a high precision experimental test measuring this temperature or magnetic field dependence would give a definitive indication of whether a diffusive model truly describes the behavior of electrons in Casimir experiments. 

Additionally, we explore the effect that fluctuations in the disorder potential can have on the Casimir energy and the validity of using the Drude model considering that the correct averaging procedure would give a result that differs from the Drude model by the inclusion of nonlocal disorder fluctuation contributions. 
We find that for a given level of disorder, one can always overcome the effects of fluctuations by holding the plates far enough apart, which justifies the use of the Drude model.

\emph{Acknowledgements} -- This work was supported by the DOE-BES (Grant No. DESC0001911) (A.A. and V.G.), the JQI-PFC (J.W.), and the Simons Foundation. 

\bibliography{references}

\end{document}